# Walking behavior in a circular arena modified by pulsed light stimulation in *Drosophila melanogaster w*$^{1118}$ line

Running title: *Drosophila* walking behavior in circular arena

Shuang Qiu [a] and Chengfeng Xiao [b,*]

[a] School of Environmental and Biological Engineering, Nanjing University of Science and Technology, Xiao Ling Wei 200, Nanjing 210094, Jiangsu, China

[b] Department of Biology, Queen's University, Kingston, Ontario K7L 3N6, Canada

* Correspondence:

C. Xiao, Department of Biology, Queen's University, Kingston, Ontario K7L 3N6, Canada.
Email: xiao.c@queensu.ca

Word count: Abstract 208; Main text 5444

# ABSTRACT

The *Drosophila melanogaster* white-eyed $w^{1118}$ line serves as a blank control, allowing genetic recombination of any gene of interest along with a readily recognizable marker. $w^{1118}$ flies display behavioral susceptibility to environmental stimulation such as light. It is of great importance to characterize the behavioral performance of $w^{1118}$ flies because this would provide a baseline from which the effect of the gene of interest could be differentiated. Little work has been performed to characterize the walking behavior in adult $w^{1118}$ flies. Here we show that pulsed light stimulation increased the regularity of walking trajectories of $w^{1118}$ flies in circular arenas. We statistically modeled the distribution of distances to center and extracted the walking structures of $w^{1118}$ flies. Pulsed light stimulation redistributed the time proportions for individual walking structures. Specifically, pulsed light stimulation reduced the episodes of crossing over the central region of the arena. An addition of four genomic copies of mini-*white*, a common marker gene for eye color, mimicked the effect of pulsed light stimulation in reducing crossing in a circular arena. The reducing effect of mini-*white* was copy-number-dependent. These findings highlight the rhythmic light stimulation-evoked modifications of walking behavior in $w^{1118}$ flies and an unexpected behavioral consequence of mini-*white* in transgenic flies carrying $w^{1118}$ isogenic background.

# INTRODUCTION

Walking behavior of *Drosophila* adults is complicated, containing rich information that is indicative of environmental impacts. For instances, flies are claustrophobic and reluctant to walk through narrow spaces (Ewing, 1961). An adult fly performs relentless walking, interspaced with several pausing, for hours in a relatively small arena (B. J. Cole, 1995; Xiao & Robertson, 2015). Individual wild-type Canton-S flies walk in one way, counter-clockwise or clockwise, for minutes in a circular arena (Xiao, Qiu, & Robertson, 2017a). Walking behavior is also largely affected by the neural function and genetic composition. Flies perform sustained backward walking if the moonwalker descending neurons in the brain are activated (Bidaye, Machacek, Wu, & Dickson, 2014). The persistence of one-way walking in the circular arena is reduced in flies carrying the $w^{1118}$ allele, a null mutation of the *white* ($w^+$) gene, in the wild-type genetic background (Xiao et al., 2017a).

The white-eyed $w^{1118}$ flies carry $w^{1118}$ allele in an isogenic background. $w^{1118}$ flies serve as a blank control, allowing transgenic recombination of a gene of interest which is often accompanied by an eye-color marker gene. $w^{1118}$ flies thus have the widest application in *Drosophila* genetics. It becomes critical to characterize the walking behavior of $w^{1118}$ flies because this task would provide an essential baseline from which the effect of an exotic gene on walking could be differentiated.

Studies have shown that $w^{1118}$ flies have increased vulnerability to the environmental stimulation. $w^{1118}$ flies display light sensitivity of locomotor recovery from anoxia in the night, whereas wild-type Canton-S flies have no such light sensitivity (Xiao & Robertson, 2017). $w^{1118}$ flies have reduced courtship behavior in the daylight (Krstic, Boll, & Noll, 2013), and increased variation of phototaxis compared with Canton-S (Kain, Stokes, & de Bivort, 2012). White-eyed flies also show reduced regularity and precision in photic orientation (W. H. Cole, 1922). Additionally, $w^{1118}$ flies have greatly reduced periodicity of episodic motor activity that has been common in wild-type, and a pulsed light stimulation increases this periodicity (Qiu, Xiao, & Robertson, 2016). These findings suggest an intrinsic feature of behavioral susceptibility to the environmental impact such as light in $w^{1118}$ flies. Therefore, we hypothesized that the walking behavior of adult $w^{1118}$ flies could be altered by pulsed light stimulation.

In the current study, we examined the walking behavior of $w^{1118}$ flies in the circular arenas and observed that pulsed light stimulation improved the regularity of 3D walking trajectories of $w^{1118}$ flies. Using combined techniques of fly tracking, behavioral computation, and genetic manipulation, we extracted the walking structures showed the specific modifications of individual structures by pulsed light stimulation. An addition of mini-*white*, a common eye-color marker gene, mimicked the effects of pulsed light stimulation in modifying the walking behavior of $w^{1118}$ flies in a manner that was copy-number-dependent.

# METHODS

## *Flies*

Flies used in this study and their sources were: isogenic line $w^{1118}$ (L. Seroude laboratory); 10×UAS-IVS-mCD8::GFP (attP2) (Bloomington *Drosophila* Stock Center (BDSC) # 32185);

10×UAS-IVS-mCD8::GFP (attP40) (BDSC # 32186); tubP-Gal80$^{ts}$ (BDSC # 7019) and NP6520-Gal4 (Awasaki, Lai, Ito, & Lee, 2008). Flies (except for tubP-Gal80$^{ts}$) were backcrossed into $w^{1118}$ flies for ten generations before testing or further genetic recombination. tubP-Gal80$^{ts}$ was generated in a *w*(CS10) background (McGuire, 2003), and thus was not tested directly. This line was used for generating a line with chromosomal recombination. Except for $w^{1118}$, all the testes lines carry an eye-color marker gene mini-*white*.

We also tested two recombination lines: 10×UAS-IVS-mCD8::GFP (attP40); 10×UAS-IVS-mCD8::GFP (attP2) and tubP-Gal80$^{ts}$ (II); NP6520-Gal4 (III) (Xiao, Qiu, & Robertson, 2017b). Multiple genomic copies of mini-*white* are included in these recombination lines. attP40 is a fixed or site-specific docking site on the second chromosome, whereas attP2 on the third (Pfeiffer et al., 2010). The insertion site for tubP-Gal80$^{ts}$ construct is on the second chromosome but not at attP40, whereas the insertion site for NP6520-Gal4 is on the third but not at attP2.

Flies were raised with standard medium (cornmeal, agar, molasses, and yeast) at 21-23 °C with 60-70 % relative humidity. A light/dark (12/12 h) cycle was provided with three light bulbs (Philips 13 W compact fluorescent energy saver) in a room around 133 square feet. Male flies were collected within 0 - 2 days after emergence. We used pure nitrogen gas to anesthetize flies during collection time. Collected flies were raised in food vials at a density of 20 - 25 flies per vial. A minimum of three days free of nitrogen exposure was guaranteed before behavioral testing. The ages of tested flies were 3 - 9 days old. To avoid natural peak activities in the mornings and evenings (Grima, Chélot, Xia, & Rouyer, 2004), experiments were performed during the light time with three hours away from the light on/off switch.

*Pulsed light stimulation*

The protocol for the delivery of pulsed light stimulation was described previously (Qiu, Xiao, & Robertson, 2016). Briefly, flies were exposed to an illumination of continuous cycles of 5 s ON - 15 s OFF of white light (Rxment® 5050 SMD LED light strip) during the 12 h daytime for the entire life cycle. The light pulse was driven by a Grass S88 stimulator (Grass Instrument Co., MA). Newly emerged flies were collected and raised at the pulsed illumination condition for four additional days free of anesthesia. Flies were allowed 1 h in the regular white light condition before testing.

*Locomotor assay*

The locomotor assay was performed by following a reported protocol (Xiao & Robertson, 2015). Briefly, flies were loaded individually into circular arenas (1.27 cm diameter and 0.3 cm depth). The depth of 0.3 cm was considered to suppress vertical movement while allowing turning around during walking. We machined 128 arenas (8 × 16) in an area of 31.0 cm × 16.0 cm Plexiglas. The bottom side of the arena was covered with thick filter paper allowing air circulation. The top was covered by a sliding Plexiglas with holes (0.3 cm diameter) at one end for fly loading. The Plexiglas with arenas was secured in a large chamber (48.0 cm × 41.5 cm × 0.6 cm). A flow of room air (2 L/min) was provided to remove the effect of dead space (Bouhuys, 1964). Illumination for the settings was provided by using a white light box (Logan portaview slide/transparency viewer) with a 5000 K ("daylight") color-corrected fluorescent lamp. Flies were allowed 5 min to adapt to the experimental settings. Walking activities were video-captured at 15 frames per second, and stored for post analysis.

*Fly tracking*

Fly positions (the locations of the center of mass) per 0.2 s were computed by a fly tracking software (Xiao & Robertson, 2015). For each fly, a dataset containing 1500 positional coordinates from 300 s locomotion was collected. These data were used for subsequent behavioral analysis, including the construction of time-series of 3D trajectory, modeling of data distribution and extraction of behavioral structures of walking.

*Construction of 3D walking trajectory*

Time-series of 3D walking trajectory was constructed by connecting the X-Y positions per 0.2 s over a period. The Cloud function from an R package lattice (Sarkar, 2008) was used for the visualization of 3D trajectory. For clarity, we present only part of the data (e.g. data from 60 s walking) in several figures.

*Modeling of data distribution*

We characterized the fly locations in the arena by modeling the observations of distances to center, a parameter reflective of location preference during walking. Fly locations were heavily distributed along the periphery but not the central region of the arena. Thus, a model distribution suitable for skewed data was preferably considered. We tested these models: Gumbel, Weibull, Gamma, Inverse Gamma, Logistic, t Family, Normal, Normal Family and Log-Normal distributions (Rigby, Stasinopoulos, Heller, & Voudouris, 2014). An ideal model should be suitable for most flies while at the same time sufficient to describe the data distributions of individual flies. The modeling was performed by following the instruction in an R package depmixS4 (Visser & Speekenbrink, 2010). There were three major steps: (1) calculate the distance to center for each sampled location; (2) fit the data with a dependent mixture model ($K = 1 - 4$); and (3) choose the best model by a classic criterion, the Bayesian information criterion (BIC) (Schwarz, 1978).

We observed that flies paused occasionally in the arenas. To avoid the overweight of fixed values of distance to center, the pausing episodes were excluded from the modeling. An episode of pausing was defined as activities containing at least five consecutive steps with step size < 0.28 mm, a threshold distance equivalent to that between two adjacent diagonal pixels.

*Extraction of walking structures*

Walking structures were the salient behavioral components or modules that were discernible due to a common feature of activities within a component. The rationale for the extraction of walking structures was based on visual characteristics of walking activities, but most importantly, the revealed features by modeling the data distribution. We observed that a three-state model, the Normal-Gumbel-Gumbel mixture, was suitable for approximating the data from most of the tested flies. Thus, data from all the tested flies were organized and fitted with this mixture model by following the instructions (Visser & Speekenbrink, 2010). The returned values of location parameter $\mu$ and posterior state for every observation were used for the classification of walking structures. We also extracted the relevant video-frames for visual validation of classification. For statistical convenience, several walking structures were defined by using a criterion - at least five consecutive locations in the same posterior state. Considering the pre-defined "pausing" episodes, we have classified five walking structures: "crossing", "side-wall walking", "angle walking", "pausing" and "unclassified".

*Statistics*

Data processing and visualization were performed by using software R (R Development Core Team, 2016) and these supplementary packages: gdata, reshape, lattice (Sarkar, 2008), ggplot2, gamlss and gamlss.mx (Rigby & Stasinopoulos, 2015), and depmixS4 (Visser & Speekenbrink, 2010). Data normality was examined by D'Agostino & Pearson omnibus normality test. Because some of the data were not normally distributed, we used nonparametric tests (i.e. Mann-Whitney test and Kruskal-Wallis test with Dunn's multiple comparisons) to examine the difference of medians between groups. Data were presented as scattered dots and median. A $P < 0.05$ was considered statistically significant.

*Ethical Note*

This study used *Drosophila melanogaster* adult flies for behavioral observation and analysis. There was a procedure for fly collection during which flies were anesthetized using pure nitrogen gas for a minimized period (less than 2 min). There was no procedure of animal dissection or tissue homogenization. Flies were exposed to regular light illumination or pulsed light stimulation with regular light intensity. All the tested flies were raised with regular culture medium at the room temperatures with regular humidity.

## RESULTS

### *Pulsed light stimulation increased the regularity of walking trajectories of $w^{1118}$ flies in circular arenas*

A four-day-old male $w^{1118}$ fly walked actively in a circular arena (1.27 cm diameter 0.3 cm depth). The 3D trajectory, represented as connected X, Y-positions over time, displayed an irregular, collapsed-coil shape. The irregular trajectories were observed consistently among individuals (Fig. 1a). Flies treated with a pulsed light stimulation (continual cycles of 5 s ON and 15 s OFF during the daytime), however, had altered 3D trajectories with visually increased regularity (Fig. 1b). 2D path showed that pulsed flies had reduced crossing over the central regions compared with controls (Fig. S1). Thus, pulsed light stimulation modified the walking behavior of $w^{1118}$ flies and increased the regularity of walking trajectories in the circular arenas.

*Distributional features for the distances to center*

To understand how pulsed light stimulation modulated the walking behavior in the arena, we statistically modeled the observations of distance to center, a parameter indicative of location preference during walking.

Datasets from all the tested flies (n = 179) in the current study were analyzed. The intermittent episodes of pausing were pre-identified and excluded from modeling. There was an uneven dispersion of fly locations in each arena. Flies had more chance located along the periphery than in the central region (Fig. S1). By fitting the data with each selected single-state model ($K = 1$), we found that the distances to center could be described primarily by two models: a Gumbel distribution for 60.9% (109/179) flies and a *t* Family distribution for 38.5% (69/179) flies (Fig. 2). Gumbel distribution was suitable for the data with negative skewness, whereas *t* Family distribution was descriptive of data with high kurtosis (Rigby et al., 2014). The other selected models (including Weibull, Gamma, Inverse-Gamma, Logistic, Normal, Normal Family and Log-Normal distributions) were disfavored, perhaps due to their suitability for the data with positive skewness, or the data with normal or low kurtosis. Therefore, the distribution of distances to center displayed co-existed features of negative skewness and high kurtosis.

Distributional features could be reflective of walking structures in a circular arena. For example, a fly would have consistently large distances to center while walking on the side-wall, and relatively small distances to center with increased variation while walking on the floor or ceiling of the arena. The potential walking structures could be approximated by a mixture model with proper components.

A two-state ($K = 2$) heterozygous model, including one Normal component and one Gumbel component, was favored to describe the datasets of 61.5% (110/179) flies. However, a two-state model homozygous for Gumbel distribution was effective for only 14.5% (26/179) flies, and a two-state *t* Family only 12.8% (23/179) (Fig. 2). A mixture of *t* Family-Gumbel (effective for 0.0% flies) was inferior to the mixture of Normal-Gumbel for modeling the distances to center. A mixture of Logistic-Gumbel (effective for 10.1% flies) was also inferior to the Normal-Gumbel model. The *t* Family, Logistic and Normal models were suitable respectively for the datasets with high kurtosis, moderate kurtosis and normal distribution (Rigby et al., 2014). Therefore, a split of kurtosis should be considered while mixing multiple components. Additionally, the approximation was disfavored with a Gumbel-Normal mixture, the inverse of Normal-Gumbel mixture, suggesting the location specificity of each component.

The remaining data after the Normal-Gumbel modeling were explainable mainly by homozygous Gumbel (for 14.5% flies), homozygous *t* Family (for 12.8% flies) and a Logistic-Gumbel mixture (for 10.1% flies). These results suggested the necessity of an additional Gumbel component to the Normal-Gumbel mixture. We thus considered a three-state ($K = 3$) model: the Normal-Gumbel-Gumbel mixture. By fitting the data with this three-state model, we observed an increased effectiveness of 81.0% (145/179) flies (Fig. 2). A similar mixture, the Logistic-Gumbel-Gumbel, however, was favored for only 17.3% (31/179) flies. The suitabilities of four-state ($K = 4$) models were also examined. We found that each of the four-state models was favored for 0.0 - 30.7% flies. Therefore, the three-state Normal-Gumbel-Gumbel model was most suitable for approximating the distribution of distances to center.

*Behavioral structures of walking in circular arena*

Data approximation provided a statistic basis for the extraction of behavioral components. We next asked whether there were model-based, identifiable walking structures common to all the tested flies in the circular arenas.

A dataset containing 237,070 observations (i.e. distances to center) from 179 flies were approximated by a Normal-Gumbel-Gumbel model. This gave rise to three values of $\mu$ (location parameter) at 5.36, 4.98 and 3.84, representing the peaks of three groups of observations (Fig. S2). The modeling also returned posterior states specific for individual observations, and a transition matrix of probabilities from state to state. We aligned the modeled states along with actual observations and extracted the relevant video-frames for visual validation. The modeled states with the lowest $\mu$ (at 3.84) were representative of locations with small distances to center (Fig. 3a). Fly positions from relevant video-frames displayed crossing activities over the central regions. Each episode of crossing last roughly 1 s, beginning with and ending at the peripheral region (Fig. 3b). We thus used the term "crossing" to denote those activities classified by the lowest $\mu$.

Two additional walking structures, "side-wall walking" and "angle walking", were identified. "Side-wall walking" was descriptive of activities with the highest $\mu$ (at 5.36), representing the walking on the distal extreme: the side wall of the arena (Fig. 3b). In these scenarios flies posed laterally to the camera. For statistical convenience, a "side-wall walking" was defined as the activities of at least five consecutive locations with the highest $\mu$. "Angle walking" was the activities of at least five consecutive locations with $\mu$ at 4.98, describing three situations: (1) flies were on the floor or ceiling, (2) flies were near the periphery of arena, and (3) body main axis often formed an acute angle with the side-wall while walking along the periphery (Fig. 3b and S3).

We added a structure of "unclassified" to represent the activities which were non-pausing, non-crossing and those failed to be categorized into "side-wall walking" or "angle walking".

Together with pre-identified pausing episodes, we categorized five walking structures: crossing, side-wall walking, angle walking, pausing and unclassified activities. Figure S3 showed the schematic walking structures with relevant fly postures and $\mu$ values.

*Pulsed light stimulation caused redistributed time proportions for walking structures*

We analyzed the time proportions for walking structures in the circular arena. Both controls and pulsed flies displayed frequent transitions among the walking structures (Fig. 4a). Time proportions for crossing in pulsed flies (median 0.09, interquartile range (IQR) 0.05 - 0.14, n = 32) were markedly smaller than those in controls (median 0.16, IQR 0.11 - 0.20, n = 32) ($P < 0.0001$, Mann-Whitney test) (Fig. 4b). However, time proportions for side-wall walking in pulsed flies (median 0.48, IQR 0.37 - 0.58, n = 32) were statistically the same as those in controls (median 0.39, IQR 0.34 - 0.49, n = 32) ($P = 0.0556$, Mann-Whitney test). Time proportions for angle walking in pulsed flies (median 0.20, IQR 0.16 - 0.28, n = 32) were greater than those in controls (median 0.17, IQR 0.13 - 0.20, n = 32) ($P = 0.0040$, Mann-Whitney test).

There was no significant difference of time proportion for pausing between pulsed flies (median 0.05, IQR 0.04 - 0.08, n = 32) and controls (median 0.07, IQR 0.04 - 0.10, n = 32) ($P$ = 0.4376, Mann-Whitney test). Time proportions for unclassified activities in pulsed flies (median 0.15, IQR 0.12 - 0.18, n = 32) was smaller than those in controls (median 0.19, IQR 0.15 - 0.23, n = 32) ($P$ = 0.0011, Mann-Whitney test) (Fig. 4b).

In summary, pulsed light stimulation reduced time proportions for crossing and unclassified activities, and increased the time proportion for angle walking, remaining the time proportions unchanged for side-wall walking and pausing. Thus, pulsed light stimulation caused redistributed time proportions for individual walking structures.

### *Effects of pulsed light stimulation on crossing, side-wall walking and angle walking*

The reduced crossing was likely the most noticeable feature in 3D trajectories in pulsed flies. We further examined the effect of pulsed light stimulation on crossing in the circular arena. The number, duration, and interval of crossing episodes were analyzed.

During 300 s locomotion, numbers of crossing episodes in pulsed flies (median 31, IQR 24 - 47, n = 32) were greatly reduced compared with those in controls (median 58, IQR 41 - 62, n = 32) ($P < 0.0001$, Mann-Whitney test) (Fig. 5a). There was no significant difference of durations of crossing between pulsed flies (median 0.8 s, IQR 0.6 - 0.8 s, n = 32) and controls (median 0.8 s, IQR 0.6 - 0.8 s, n = 32) ($P$ = 0.7023, Mann-Whitney test). Intervals of crossing in pulsed flies (median 4.7 s, IQR 3.0 - 7.8 s, n = 32) were greatly increased compared with those in controls (median 2.7 s, IQR 2.2 - 3.7 s, n = 32) ($P < 0.0001$, Mann-Whitney test). Thus, pulsed light stimulation led to a decreased number of crosses and an increased interval, remaining the duration unaffected.

Both side-wall walking and angle walking took relatively large time proportions during walking. We examined the effects of pulsed light stimulation on side-wall walking and angle walking. Numbers, durations, and intervals of side-wall walking in pulsed flies were statistically the same as those in controls (Fig. 5b). Numbers of angle walking episodes in pulsed flies were comparable with those in controls ($P$ = 0.2015, Mann-Whitney test). However, durations of angle walking in pulsed flies (median 1.4 s, IQR 1.4 - 1.6 s, n = 32) were longer than those in controls (median 1.3 s, IQR 1.2 - 1.4 s, n = 32) ($P$ = 0.0005, Mann-Whitney test). There was no significant difference of intervals of angle walking episodes between pulsed flies and controls ($P$ = 0.053, Mann-Whitney test) (Fig. 5c). Therefore, pulsed light stimulation caused the increased duration of angle walking episodes.

### *The mini-white gene reduced crossing in circular arena*

$w^{1118}$ flies carry the $w^{1118}$ allele, a null mutation of $w^+$, in an isogenic chromosomal background. $w^+$ has a well-known responsibility for eye color. Additionally, it has been proposed that $w^+$ possesses housekeeping functions (Xiao & Robertson, 2016; Xiao & Robertson, 2017). We examined whether the addition of mini-*white* (m$w^+$), a miniature form of $w^+$ and also a common eye-color marker gene, reduced the crossing of $w^{1118}$ flies in the circular arenas. There is of great significance for this examination because it was unwanted if m$w^+$ had a role to modify the walking behavior except for the desired role for eye color. We chose UAS transgenic flies with m$w^+$ inserted into site-specific docking sites (i.e. attP2 and attp40) in the autosomes (Pfeiffer et al., 2010). Selected flies were backcrossed into $w^{1118}$ flies for ten generations to have synchronized isogenic background.

It was observed that the numbers of crosses were reduced in flies carrying four genomic copies of m$w^+$ (appeared as two homozygous alleles in the line 10×UAS-IVS-mCD8::GFP(attP40); 10×UAS-IVS-mCD8::GFP(attP2)) compared with flies carrying two genomic copies of m$w^+$ (two heterozygous alleles in 10×UAS-IVS-mCD8::GFP(attP40)/+; 10×UAS- IVS-mCD8::GFP (attP2)/+) (Fig. 6a).

Flies carrying one genomic copy of m$w^+$ (10×UAS-IVS-mCD8::GFP (attP2)/+) had the numbers of crosses (median 42.0, IQR 34.5 - 51.0, n = 14) statistically the same as $w^{1118}$ flies ($P > 0.05$, Kruskal-Wallis test with Dunn's multiple comparisons) (Fig. 6b). Another fly line carrying one genomic copy of m$w^+$ (10×UAS-IVS-mCD8::GFP(attP40)/+) also had the numbers of crosses (median 58.0, IQR 38.8 - 64.5, n = 16) statistically the same as $w^{1118}$ flies ($P > 0.05$, Kruskal-Wallis test with Dunn's multiple comparisons) (Fig. 6b).

Effect of two genomic copies of m$w^+$ on the number of crosses was examined. Within 300 s walking in the circular arena, flies carrying two genomic copies of m$w^+$ (homozygous 10×UAS-IVS-mCD8::GFP (attP2)) had the numbers of crosses (median 43.0, IQR 31.0 – 53.0, n = 20) the same as $w^{1118}$ flies ($P > 0.05$, Kruskal-Wallis test with Dunn's multiple comparisons). However, flies carrying two genomic copies of m$w^+$ (homozygous 10×UAS-IVS-mCD8::GFP (attP40)) had the numbers of crosses (median 36.0, IQR 18.5 - 47.5, n = 8) lower than $w^{1118}$ flies ($P < 0.05$, Kruskal-Wallis test with Dunn's multiple comparisons). Flies carrying two heterozygous m$w^+$ alleles (10×UAS-IVS-mCD8::GFP(attP40)/+; 10×UAS-IVS-mCD8::GFP (attP2)/+) showed the numbers of crosses (median 46.5, IQR 38.3 - 56.8, n = 16) comparable with $w^{1118}$ flies ($P > 0.05$, Kruskal-Wallis test with Dunn's multiple comparisons). Flies carrying two heterozygous m$w^+$ alleles at non-fixed docking sites (tub-Gal80$^{ts}$(II)/+; NP6520-Gal4(III)/+) had the numbers of crosses (median 54.0, IQR 50.0 - 65.3, n = 16) also comparable with $w^{1118}$ flies ($P > 0.05$, Kruskal-Wallis test with Dunn's multiple comparisons).

During 300 s walking, flies with four genomic copies of m$w^+$ (homozygous 10×UAS-IVS-mCD8::GFP(attP40); 10×UAS-IVS-mCD8::GFP(attP2)) had the numbers of crosses (median 17.0, IQR 6.0 - 27.0, n = 11) greatly reduced compared with $w^{1118}$ flies ($P < 0.001$, Kruskal-Wallis test with Dunn's multiple comparisons). Another fly line carrying four genomic copies of m$w^+$ in non-fixed docking sites (homozygous tub-Gal80$^{ts}$(II); NP6520-Gal4(III)) had the numbers of crosses (median 27.0, IQR 17.8 - 32.8, n = 14) also markedly reduced compared with $w^{1118}$ flies ($P < 0.001$, Kruskal- Wallis test with Dunn's multiple comparisons).

Taken together, four genomic copies of m$w^+$ reduced the number of crosses during walking, whereas one or two genomic copies of m$w^+$ had an inclusive effect. Thus, four genomic copies of m$w^+$ mimicked the effect of pulsed light stimulation in reducing the crossing in the circular arena.

***Copy-number-dependent effect of mini-*white** *on crossing*

The numbers of crosses in the circular arena in flies with four genomic copies of m$w^+$ (homozygous 10×UAS-IVS- mCD8::GFP(attP40);10×UAS-IVS-mCD8::GFP(attP2)) were significantly lower than those in flies with two genomic copies of m$w^+$ (10×UAS-IVS-mCD8::GFP(attP40)/+;10×UAS-IVS-mCD8::GFP(attP2)/+) ($P < 0.01$, Kruskal-Wallis test with Dunn's multiple comparisons) (Fig. 6b). Both fly lines had the consistent genomic background. Thus, the effect of m$w^+$ on the number of crosses was copy-number-dependent. Similarly, the numbers of crosses in flies with four genomic copies of m$w^+$ (homozygous tub-Gal80$^{ts}$(II); NP6520-Gal4(III)) were lower than those in flies with two genomic copies of m$w^+$ (tub-Gal80$^{ts}$(II)/+; NP6520-Gal4(III)/+) ($P < 0.001$, Kruskal-Wallis test with Dunn's multiple comparisons). These data supported the copy-number-dependent effect of m$w^+$ in reducing the number of crosses in the arena.

## DISCUSSION

The *Drosophila* $w^{1118}$ line is susceptible to light illumination (Krstic et al., 2013; Qiu et al., 2016; Xiao & Robertson, 2017). In this study, we show that pulsed light stimulation modifies the walking behavior of $w^{1118}$ flies in circular arenas. Specifically, pulsed light stimulation reduces the episodes of crossing over the central region of the arena. An addition of m$w^+$ mimics the effects of pulsed light stimulation in reducing crossing in a manner that is copy-number-dependent. The current work has three layers of significance.

### *Walking structures in a circular arena*

Characterization of walking behavior of $w^{1118}$ flies provides a baseline from which the effect of a transgene could be differentiated. It is the $w^{1118}$ line but not wild-type that is often used for transgenic recombination because of two advantages. The first is the isogenic background which carries no known mutation except for the $w^{1118}$ allele. The second is the white-eyed background that allows genetic recombination accompanied with an eye-color marker gene.

The distances to center of a fly in the arena could be statistically separated into three groups, each having a specific peak (or location parameter). The group with the largest peak value is representative of fly positions on the side-wall of the arena and is related to the walking structure "side-wall walking". The side-wall is the distal extreme to the center. Due to the spatial limit, a fly would have mostly similar distances to center while walking on the side-wall. This could explain the best approximation by a Normal distribution. Two other groups have relatively small peak values, and are related to "angle walking" and "crossing". This classification is considered reasonable because flies on the floor or ceiling (but not on the side-wall) have increased freedom to change locations, either walking along the peripheral region or crossing over the central part of the arena. Most importantly, these two groups of data are negatively skewed due to the best-fit of Gumbel but not Normal distribution. The results indicate a preference of flies for the peripheral region of arenas during walking. Our finding is consistent with the observation that $w^{1118}$ flies have boundary preference in open-field arenas (Liu, Davis, & Roman, 2007; Soibam et al., 2012).

The distances to center of each fly are sequentially organized into a time-series. The walking structures, classified by grouping the sequences of time series, are in a secondary organization that would inherit the sequential information. Such a secondary organization of walking structures could be specific to the status of flies impacted by their environment. Pulsed light stimulation causes re-allotted time proportions for individual walking structures. More interestingly, pulsed light stimulation has differential effects on individual walking structures. Pulsed light stimulation decreases the number of crosses and increases the interval, remaining the duration of crossing unchanged. However, pulsed light stimulation has no effect on the number, interval or duration of side-wall walking. Sequential organization of walking structures could also be specific to the genotype of flies. By adding four genomic copies of mw+ to $w^{1118}$ flies, the episodes of crossing are greatly reduced. These findings support the notion that the sequential organization of walking structures reflects the environmental impact or genetic composition of a fly. Therefore, the behavioral consequence of a transgene, if inserted into the $w^{1118}$ background, could be deduced from the difference between the transgenic flies and $w^{1118}$ flies.

### *Behavioral modification by pulsed light stimulation*

There is an apparent clinical relevance that rhythmic external stimulation induces a behavioral effect. Rhythmic stimulation by transcranial application of oscillating potentials (0.75 Hz) induces slow oscillation-like potential fields and enhances memory retention in healthy humans (Marshall, Helgadóttir, Mölle, & Born, 2006). Repetitive transcranial magnetic stimulation (at 10 Hz) delivered to ipsilateral posterior parietal cortex increases visual short-term memory capacity (Sauseng et al., 2009). The behavioral effect of pulsed light stimulation in *Drosophila* supports the application of rhythmic external stimulation for the improvement of motor ability and cognition. Moreover, this work would expand the dimension for fine-tuning the frequency or spectrum of light stimulation, as well as the genetic basis underlying the stimulation-induced behavioral modification.

The reason to use pulsed light stimulation (with continuous cycles of 5s ON 15s OFF) is that such a rhythmicity is associated with an intrinsic episodic motor activity in wild-type Canton-S flies (Qiu et al., 2016). We have previously shown that pulsed light stimulation induces increased episodic motor activities and increased boundary preference in $w^{1118}$ flies (Qiu et al., 2016). In addition, Canton-S flies barely cross over the central regions of arenas and primarily walk in one direction, either counter-clockwise or clockwise, along with the side-walls in the circular arenas (Xiao et al., 2017a). A reduction of crossing in the arena implies a shift of behavioral performance of $w^{1118}$ flies towards wild-type. Therefore, the modifications by pulsed light stimulation suggest a restoration of behavioral and neural functions in $w^{1118}$ flies. The major genetic changes in $w^{1118}$ flies are the null mutation of $w^+$ and the isogenic chromosomal background as compared with wild-type background. These genetic changes might be associated with altered electrophysiological or behavioral performance.

### *Effect of mw+ on walking behavior*

The eye-color marker gene m$w^+$ is commonly introduced into the genome of $w^{1118}$ flies during transgenic recombination. It is anticipated that m$w^+$ would not have any neural or behavioral effect, except for its responsibility for eye color. We observed that m$w^+$ reduced crossing in the circular arena and that this effect was m$w^+$ copy-number-dependent. We have previously shown that m$w^+$ promotes persistent one-way walking in the circular arena in a manner that is copy-number-dependent (Xiao et al., 2017a). These studies support a clear role for m$w^+$ in behavioral modification.

m$w^+$ is a miniature form of $w^+$, a classic eye-color gene in *Drosophila* (Morgan, 1910). $w^+$ encodes a hemi-unit of an ATP-binding cassette (ABC) transporter, which loads cellular vesicles or granules with neurotramsmiters, biogenic amines, second messenger, pigment precursors and many small substrates for intracellular trafficking (Anaka et al., 2008; Borycz, Borycz, Kubow, Lloyd, & Meinertzhagen, 2008; Evans, Day, Cabrero, Dow, & Davies, 2008; O'Hare, Murphy, Levis, & Rubin, 1984; Sitaraman et al., 2008; D T Sullivan & Sullivan, 1975; David T. Sullivan, Bell, Paton, & Sullivan, 1979). It has been observed that $w^+$ has extra-retinal neural functions (Campbell & Nash, 2001; Hing & Carlson, 1996; Sitaraman et al., 2008; Xiao et al., 2017a; Xiao & Robertson, 2016; Zhang et al., 1995). This pleiotropic effect has led to a proposal that $w^+$ is a housekeeping gene mainly expressed in the central nervous system (Xiao et al., 2017a, 2017b). The proposed housekeeping function of m$w^+$ might be essential for reducing crossing of $w^{1118}$ flies in circular arenas.

The effect of m$w^+$ on walking behavior is unwanted. We show that flies with behavioral modifications carry one pair of homozygous m$w^+$ alleles at the site of attP40 (in the second chromosome) and another pair at attP2 (in the third chromosome). Both attP40 and attP2 are the site-specific docking sites for genetic recombination (Markstein, Pitsouli, Villalta, Celniker, & Perrimon, 2008). *Drosophila* split-Gal4 lines (~ 7,300 lines so far) carry the same docking sites for the integration of two DNA constructs: one for the Gal4 activating domain (AD) and another for DNA-binding domain (DBD) (Luan, Peabody, Vinson, & White, 2006). m$w^+$ has been included in each construct. A functional split-Gal4 line contains both constructs at attP40 and attP2. Such a genetic manipulation, however, introduces multiple genomic copies of m$w^+$ into the genome, increasing the possibility to cause behavioral or neural consequences. The copy-number-dependent effect of m$w^+$ on walking behavior supports this possibility. These findings highlight a behavioral consequence potentially attributable to m$w^+$ in transgenic flies with $w^{1118}$ isogenic background.

## ACKNOWLEDGEMENTS

This work was supported by a Startup Foundation of Nanjing University of Science and Technology (to S.Q.), and Fundamental Research Funds for the Central Universities from Nanjing University of Science and Technology (AE91319/020).

## REFERENCES

Anaka, M., MacDonald, C. D., Barkova, E., Simon, K., Rostom, R., Godoy, R. A., … Lloyd, V. (2008). The *white* gene of *Drosophila melanogaster* encodes a protein with a role in courtship behavior. *Journal of Neurogenetics*, *22*(4), 243–276. https://doi.org/10.1080/01677060802309629

Awasaki, T., Lai, S.-L., Ito, K., & Lee, T. (2008). Organization and postembryonic development of glial cells in the adult central brain of *Drosophila*. *Journal of Neuroscience*, *28*(51),

13742–13753. https://doi.org/10.1523/JNEUROSCI.4844-08.2008

Bidaye, S. S., Machacek, C., Wu, Y., & Dickson, B. J. (2014). Neuronal control of *Drosophila* walking direction. *Science*, *344*(6179), 97–101. https://doi.org/10.1126/science.1249964

Borycz, J., Borycz, J. A., Kubow, A., Lloyd, V., & Meinertzhagen, I. A. (2008). *Drosophila* ABC transporter mutants *white*, *brown* and *scarlet* have altered contents and distribution of biogenic amines in the brain. *Journal of Experimental Biology*, *211*(21), 3454–3466. https://doi.org/10.1242/jeb.021162

Bouhuys, A. (1964). Respiratory dead space. Section 3: Respiration. *Handbook of Physiology*, *1*, 699.

Campbell, J. L., & Nash, H. A. (2001). Volatile general anesthetics reveal a neurobiological role for the *white* and *brown* genes of *Drosophila melanogaster*. *Journal of Neurobiology*, *49*(4), 339–349. https://doi.org/10.1002/neu.10009

Cole, B. J. (1995). Fractal time in animal behavior: the moment activity of *Drosophila*. *Animal Behaviour*, *50*(5), 1317–1324. https://doi.org/10.1016/0003-3472(95)80047-6

Cole, W. H. (1922). Note on the relation between the photic stimulus and the rate of locomotion in *Drosophila*. *Science*, *55*(1434), 678–679.

Evans, J. M., Day, J. P., Cabrero, P., Dow, J. A. T., & Davies, S.-A. (2008). A new role for a classical gene: White transports cyclic GMP. *Journal of Experimental Biology*, *211*(6), 890–899. https://doi.org/10.1242/jeb.014837

Ewing, A. W. (1961). Attempts to select for spontaneous activity in *Drosophila melanogaster*. *Animal Behaviour*, *11*(2–3), 369–378. https://doi.org/10.1016/S0003-3472(63)80128-1

Grima, B., Chélot, E., Xia, R., & Rouyer, F. (2004). Morning and evening peaks of activity rely on different clock neurons of the *Drosophila* brain. *Nature*, *431*(7010), 869–873. https://doi.org/10.1038/nature02935

Hing, A. L., & Carlson, J. R. (1996). Male-male courtship behavior induced by ectopic expression of the *Drosophila white* gene: role of sensory function and age. *J Neurobiol*, *30*(4), 454–464. https://doi.org/10.1002/(SICI)1097-4695(199608)30:4<454::AID-NEU2>3.0.CO;2-2 [pii]\r10.1002/(SICI)1097-4695(199608)30:4<454::AID-NEU2>3.0.CO;2-2

Kain, J. S., Stokes, C., & de Bivort, B. L. (2012). Phototactic personality in fruit flies and its suppression by serotonin and white. *Proceedings of the National Academy of Sciences*, *109*(48), 19834–19839. https://doi.org/10.1073/pnas.1211988109

Krstic, D., Boll, W., & Noll, M. (2013). Influence of the *white* locus on the courtship behavior of *Drosophila* males. *PLoS ONE*, *8*(10), e77904. https://doi.org/10.1371/journal.pone.0077904

Liu, L., Davis, R. L., & Roman, G. (2007). Exploratory activity in *Drosophila* requires the *kurtz* nonvisual arrestin. *Genetics*, *175*(3), 1197–1212. https://doi.org/10.1534/genetics.106.068411

Luan, H., Peabody, N. C., Vinson, C. R., & White, B. H. (2006). Refined spatial manipulation of neuronal Function by combinatorial restriction of transgene expression. *Neuron*, *52*(3), 425–436. https://doi.org/10.1016/j.neuron.2006.08.028

Markstein, M., Pitsouli, C., Villalta, C., Celniker, S. E., & Perrimon, N. (2008). Exploiting position effects and the gypsy retrovirus insulator to engineer precisely expressed transgenes. *Nature Genetics*, *40*(4), 476–483. https://doi.org/10.1038/ng.101

Marshall, L., Helgadóttir, H., Mölle, M., & Born, J. (2006). Boosting slow oscillations during sleep potentiates memory. *Nature*, *444*(7119), 610–613. https://doi.org/10.1038/nature05278

McGuire, S. E. (2003). Spatiotemporal rescue of memory dysfunction in *Drosophila*. *Science*, *302*(5651), 1765–1768. https://doi.org/10.1126/science.1089035

Morgan, T. H. (1910). Sex-limited inheritance in *Drosophila*. *Science*, *32*, 120–122. https://doi.org/10.1126/science.32.812.120

O'Hare, K., Murphy, C., Levis, R., & Rubin, G. M. (1984). DNA sequence of the *white* locus of *Drosophila melanogaster*. *Journal of Molecular Biology*, *180*(3), 437–455. https://doi.org/10.1016/0022-2836(84)90021-4

Pfeiffer, B. D., Ngo, T. T. B., Hibbard, K. L., Murphy, C., Jenett, A., Truman, J. W., & Rubin, G. M. (2010). Refinement of tools for targeted gene expression in *Drosophila*. *Genetics*, *186*(2), 735–755. Retrieved from http://www.genetics.org/cgi/doi/10.1534/genetics.110.119917%5Cnpapers2://publication/doi/10.1534/genetics.110.119917

Qiu, S., Xiao, C., & Robertson, R. M. (2016). Pulsed light stimulation increases boundary preference and periodicity of episodic motor activity in *Drosophila melanogaster*. *PLoS ONE*, *11*(9), e0163976. https://doi.org/10.1371/journal.pone.0163976

R Development Core Team. (2016). R: A Language and Environment for Statistical Computing. *R Foundation for Statistical Computing Vienna Austria*, *0*, {ISBN} 3-900051-07-0. https://doi.org/10.1038/sj.hdy.6800737

Rigby, B., & Stasinopoulos, M. (2015). The generalized additive models for location, scale and shape in R. *J Stat Softw*, *54*(3), 1–2.

Rigby, B., Stasinopoulos, M., Heller, G., & Voudouris, V. (2014). The distribution toolbox of GAMLSS. *The GAMLSS Team*.

Sarkar, D. (2008). Multivariate Data Visualization with R. *Use R!* https://doi.org/10.1007/978-0-387-75969-2

Sauseng, P., Klimesch, W., Heise, K. F., Gruber, W. R., Holz, E., Karim, A. A., … Hummel, F. C. (2009). Brain oscillatory substrates of visual short-term memory capacity. *Current Biology*, *19*(21), 1846–1852. https://doi.org/10.1016/j.cub.2009.08.062

Schwarz, G. (1978). Estimating the dimension of a Model. *The Annals of Statistics*, *6*(2), 461–464. https://doi.org/10.1214/aos/1176344136

Sitaraman, D., Zars, M., Laferriere, H., Chen, Y.-C., Sable-Smith, A., Kitamoto, T., … Zars, T.

(2008). Serotonin is necessary for place memory in *Drosophila*. *Proceedings of the National Academy of Sciences of the United States of America*, *105*(14), 5579–84. https://doi.org/10.1073/pnas.0710168105

Soibam, B., Mann, M., Liu, L., Tran, J., Lobaina, M., Kang, Y. Y., … Roman, G. (2012). Open-field arena boundary is a primary object of exploration for *Drosophila*. *Brain and Behavior*, *2*(2), 97–108. https://doi.org/10.1002/brb3.36

Sullivan, D. T., Bell, L. A., Paton, D. R., & Sullivan, M. C. (1979). Purine transport by Malpighian tubules of pteridine-deficient eye color mutants of *Drosophila melanogaster*. *Biochemical Genetics*, *17*(5–6), 565–573. https://doi.org/10.1007/BF00498891

Sullivan, D. T., & Sullivan, M. C. (1975). Transport defects as the physiological basis for eye color mutants of *Drosophila melanogaster*. *Biochemical Genetics*, *13*(9–10), 603–613. https://doi.org/10.1007/BF00484918

Visser, I., & Speekenbrink, M. (2010). depmixS4: An R-package for hidden Markov models. *Journal of Statistical Software*, *36*(7), 1–21.

Xiao, C., Qiu, S., & Robertson, R. M. (2017a). Persistent one-way walking in a circular arena in *Drosophila melanogaster* Canton-S strain. *Behavior Genetics*, 1–14. https://doi.org/10.1007/s10519-017-9881-z

Xiao, C., Qiu, S., & Robertson, R. M. (2017b). The *white* gene controls copulation success in *Drosophila melanogaster*. *Scientific Reports*, *7*(1), 7712. https://doi.org/10.1038/s41598-017-08155-y

Xiao, C., & Robertson, R. M. (2015). Locomotion induced by spatial restriction in adult Drosophila. *PLoS ONE*, *10*(9), e0135825.

Xiao, C., & Robertson, R. M. (2016). Timing of locomotor recovery from anoxia modulated by the *white* gene in *Drosophila*. *Genetics*, *203*(2), 787–797. https://doi.org/10.1534/genetics.115.185066

Xiao, C., & Robertson, R. M. (2017). White -cGMP interaction promotes fast locomotor recovery from anoxia in adult *Drosophila*. *PLoS ONE*, *12*(1). https://doi.org/10.1371/journal.pone.0168361

Zhang, S. D., Odenwald, W. F., Odenwald, F., Zhang, S. D., Odenwald, W. F., Odenwald, F., … Odenwald, W. F. (1995). Misexpression of the *white* (*w*) gene triggers male-male courtship in *Drosophila*. *Proceedings of the National Academy of Sciences of the United States of America*, *92*(12), 5525–5529. https://doi.org/10.1073/pnas.92.12.5525

**FIGURE LEGENDS**

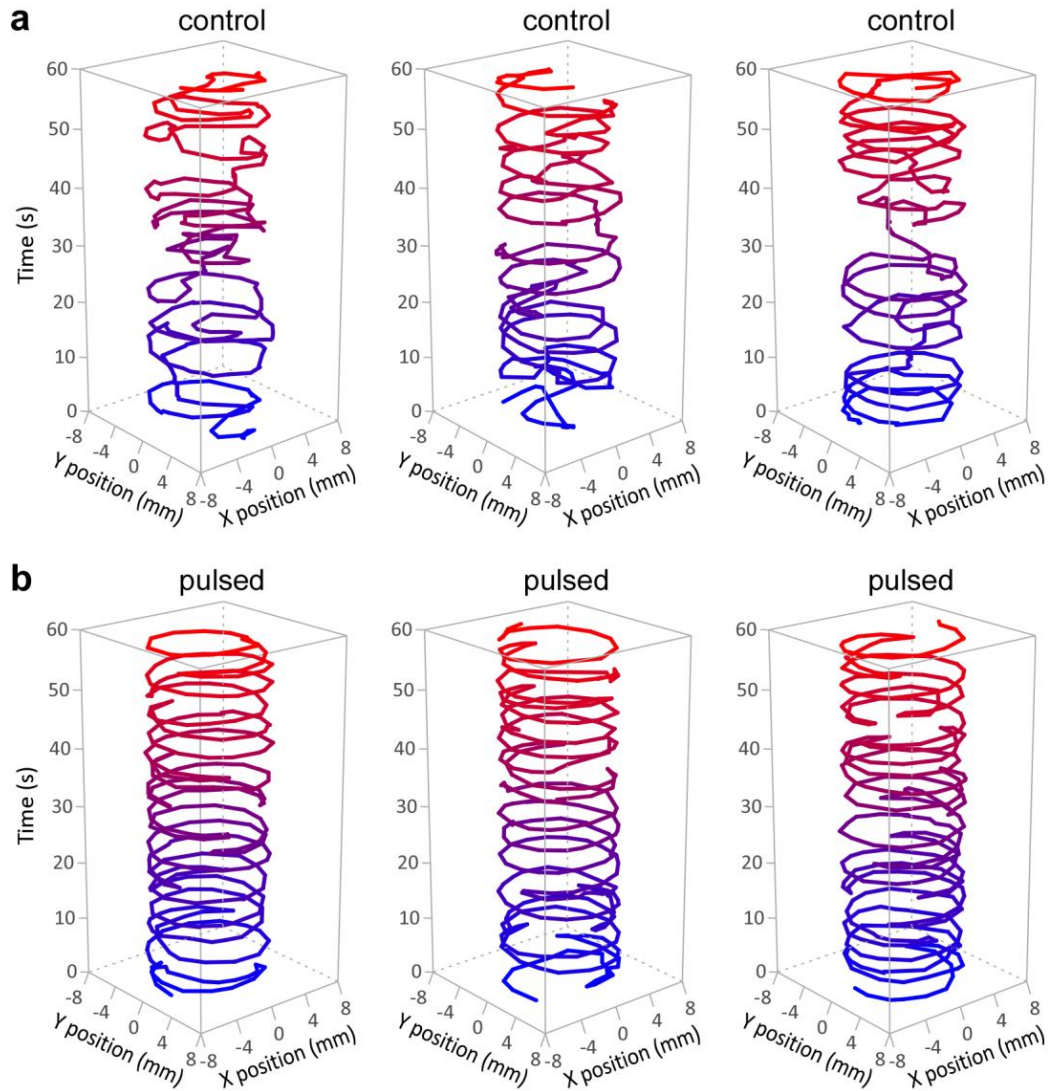

Figure 1: Pulsed light stimulation increased the regularity of 3D walking trajectories of $w^{1118}$ flies. (**a**) 3D trajectories of controls in circular arenas. (**b**) 3D trajectories of pulsed flies in circular arenas. Individual adult flies were loaded into each circular arena (1.27 cm diameter 0.3 cm depth). Fly positions were tracked once every 0.2 s. 3D trajectories were re-constructed as connected time-series of x, y-coordinates over time. For clarity, only 60 s walking from each fly was shown. Color gradients (from blue to red) indicate the time progression.

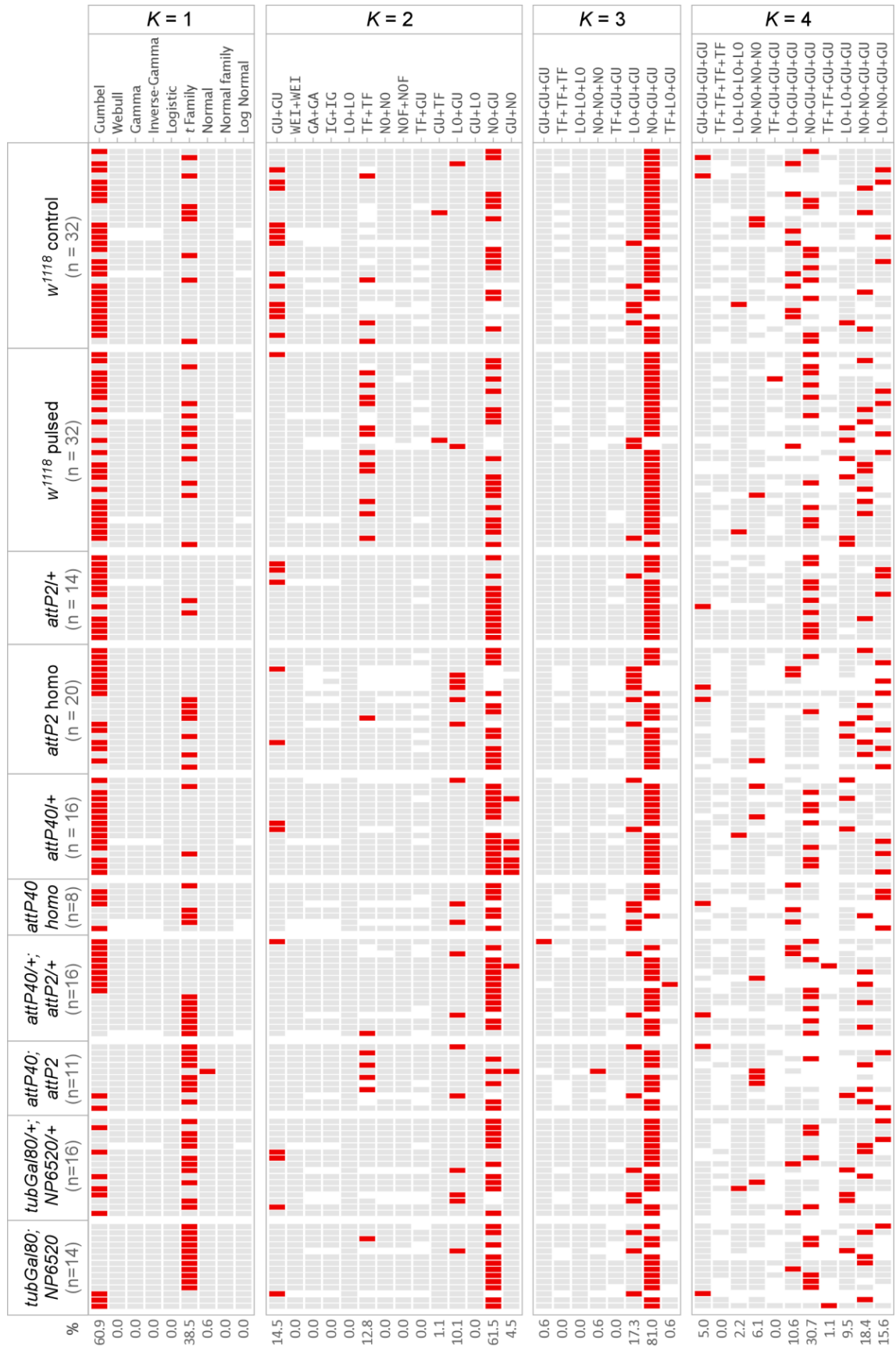18

Figure 2: Modeling the distribution of distances to center. The datasets of 179 flies were fitted with the selected models by following an instruction (Visser & Speekenbrink, 2010). Models containing single or multiple components ($K$ = 1 - 4) were applied. The Bayesian information criterion (BIC) was used for model selection. These models were tested: Gumbel (GU), Weibull (WEI), Gamma (GA), Inverse-Gamma (IG), Logistic (LO), $t$ Family (TF), Normal (NO), Normal Family (NOF) and Log-Normal distributions. Fly genotypes and sample sizes were indicated. The percent suitability was provided. Notes: red, the best-fit model (associated with the lowest BIC); grey, candidate model with greater BIC; empty space, model not applicable.

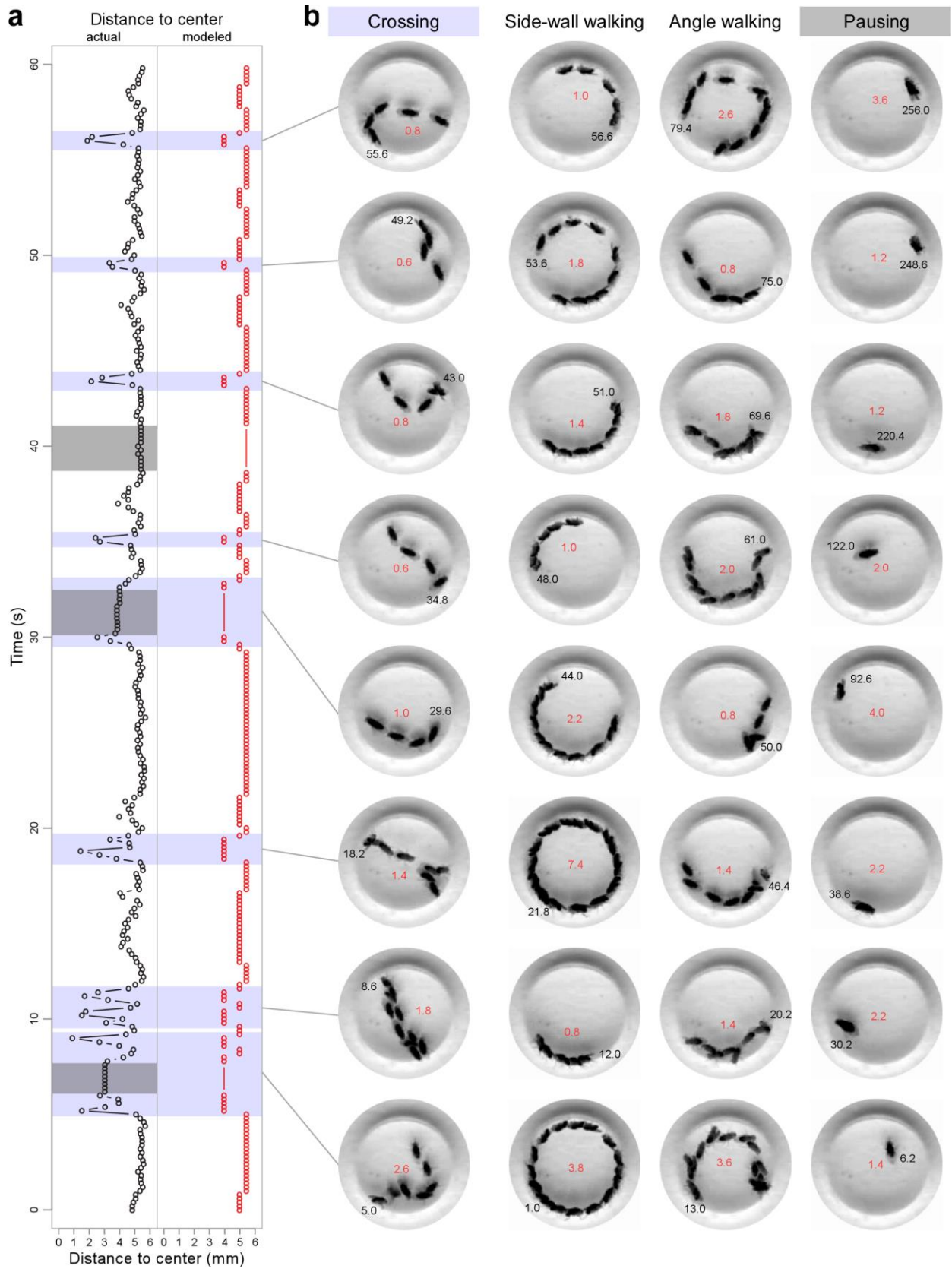

Figure 3: Classification of walking structures of $w^{1118}$ flies in circular arenas. (**a**) The alignment of actual distances to center (black) and modeled values (red) during 60 s locomotion. Blue-shaded data are associated with the structure "crossing", and grey-shaded pausing. (**b**) Visual proofs for the following walking structures: crossing, side-wall walking, angle walking and pausing. There is only one fly in each arena. Multiple relocations are from overlapped video-frames. Black number in each image indicates the starting time (s) and red number duration (s).

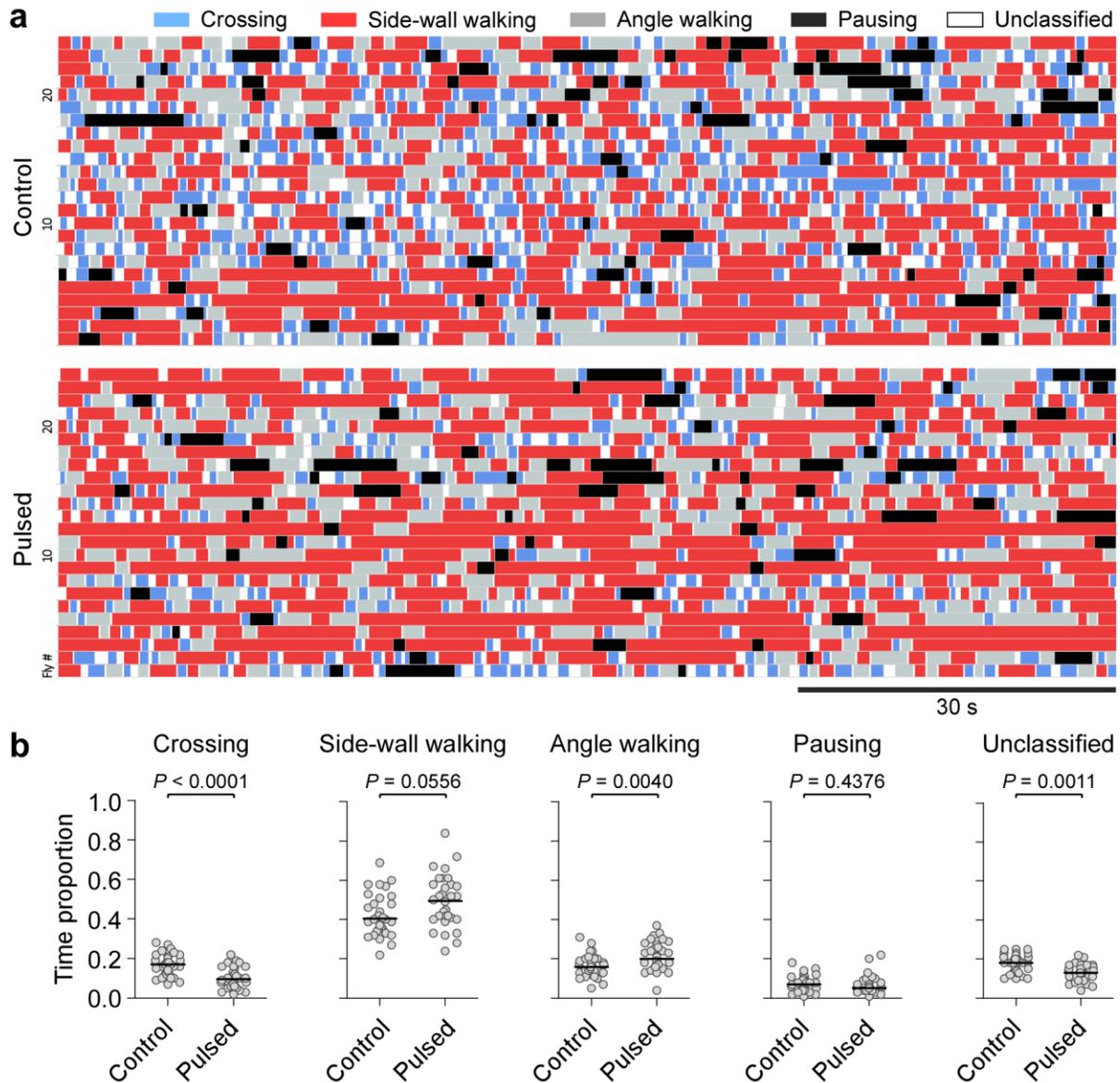

Figure 4: Walking structures of $w^{1118}$ flies in circular arenas. (**a**) Schematic representations of walking structures in a circular arena. The walking structures are: crossing (blue), side-wall walking (red), angle walking (grey), pausing (black) and unclassified activities (white). Each horizontal ribbon denotes the sequential organization of walking structures from a single fly. Walking structures from 24 controls and 24 pulsed flies are shown. Bar, 30 s. (**b**) Time proportions for individual walking structures in controls and pulsed flies during 300 s walking. Data are presented as scattered dots and median (black line). *P* values from Mann-Whitney test.

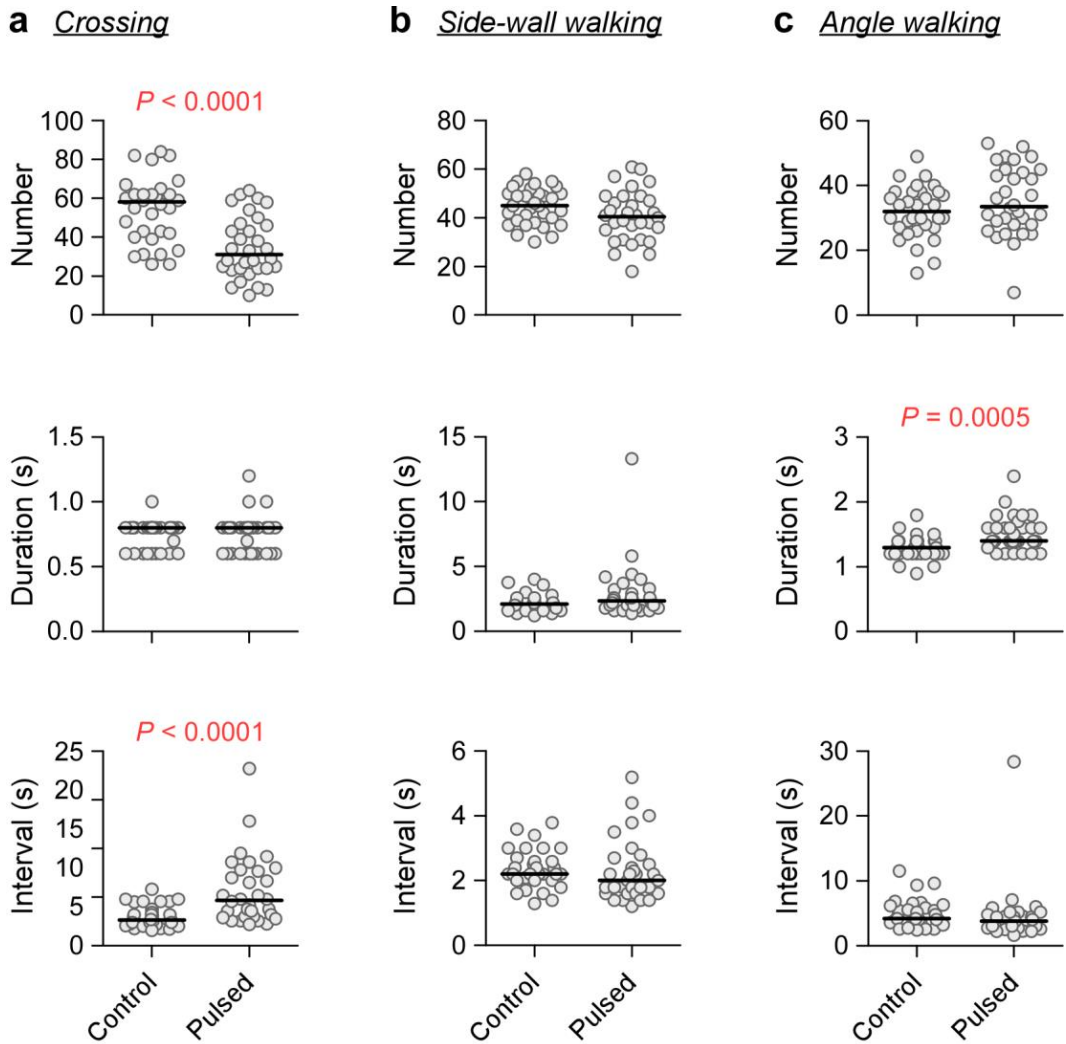

Figure 5: Pulsed light stimulation modified walking structures of $w^{1118}$ flies in circular arenas. Behavioral modifications on crossing (**a**), side-wall walking (**b**), and angle walking (**c**) by pulsed light stimulation. Data are presented as scattered dots and median (black line). *P* values from Mann-Whitney test.

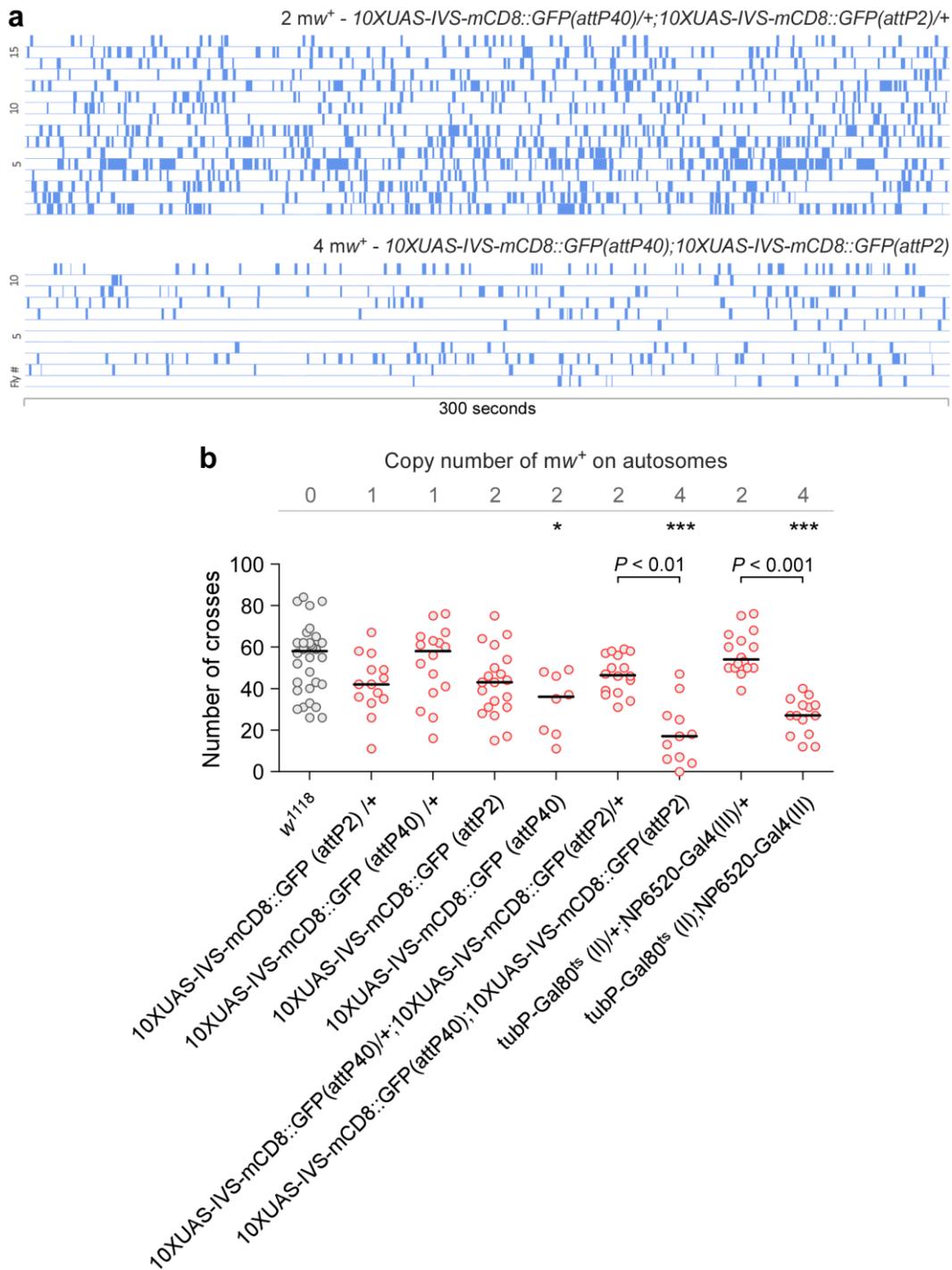

Figure 6: Four genomic copies of m$w^+$ reduced crossing in a circular arena. (**a**) Crossing episodes (blue blocks) in flies carrying two or four genomic copies of m$w^+$. The genomic copies of m$w^+$ and genotypes are indicated. Bar, 300 s. (**b**) Four genomic copies of m$w^+$ reduced the number of crosses in a circular arena. Tested flies carry 0 - 4 genomic copies of m$w^+$ in the autosomes. Data of $w^{1118}$ flies (grey) are duplicated from Figure 5 for comparison. Data of m$w^+$-carrying flies are marked red. * ($P < 0.05$, compared with $w^{1118}$), *** ($P < 0.001$, compared with w1118) and *P* values are from Kruskal-Wallis test with Dunn's multiple comparisons.

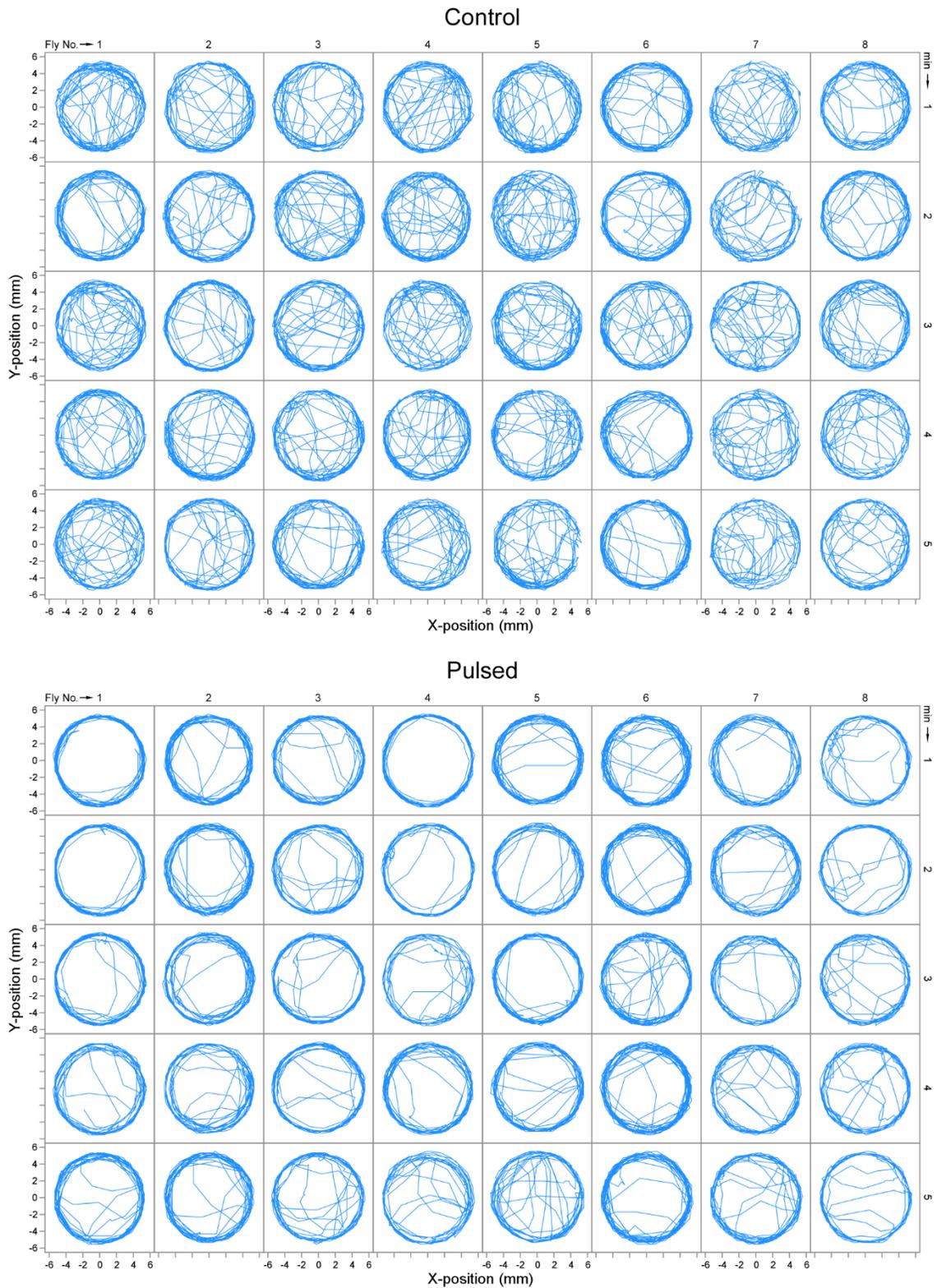

Figure S1: Walking activities of $w^{1118}$ flies in circular arenas. 2D view of walking activities of control flies (upper panel) or pulsed flies (lower panel) in circular arenas. Each panel illustrates the activities of 8 flies by 5 min. Individual flies were loaded into each arena.

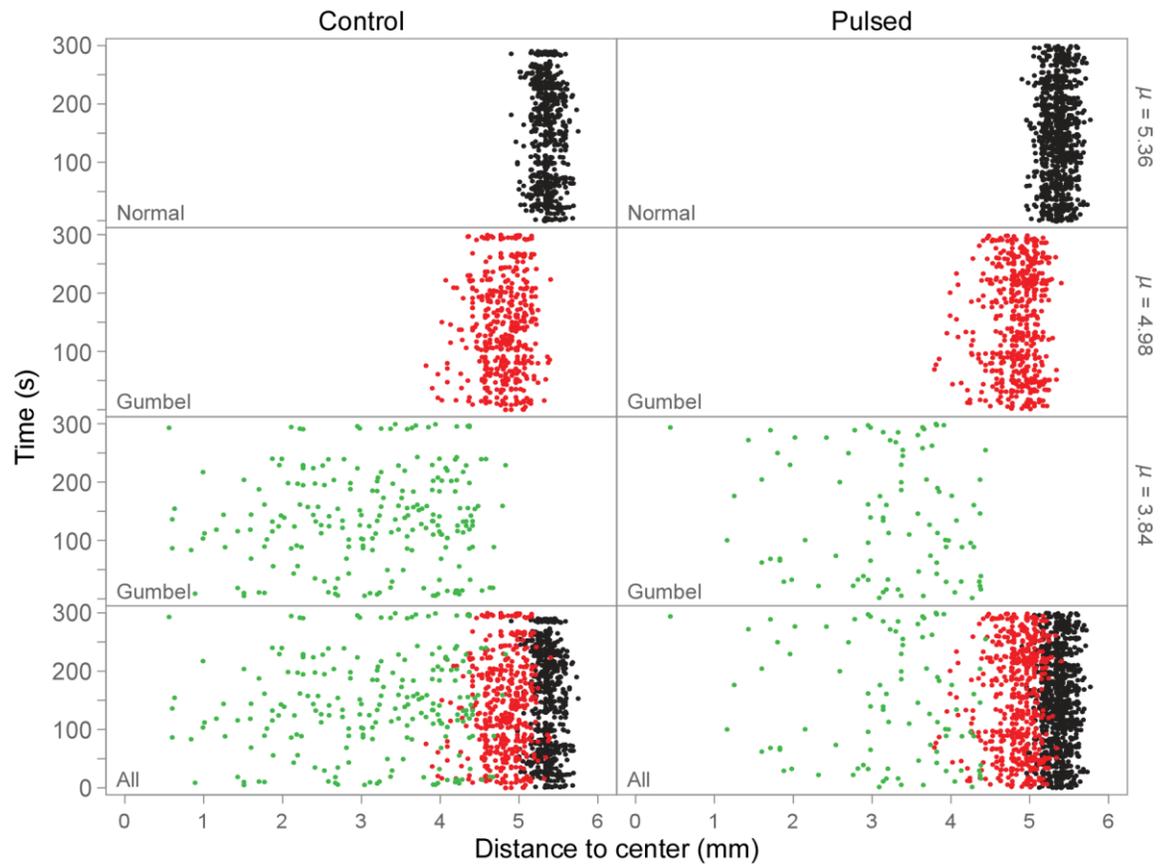

Figure S2: Three groups of distances to center separated by modeling. The distribution of distances to center was modeled by a dependent mixture model (the Normal-Gumbel-Gumbel model). The returned posterior states were used for data separation. Shown are the data from one control and one pulsed flies during 300 s walking. The distances to center are separated into three groups: a group (with $\mu$ at 5.36, black) that follows a Normal distribution; a group (with $\mu$ at 4.98, red) that displays a Gumbel distribution; and a group (with $\mu$ at 3.84, green) that also displays a Gumbel distribution. The bottom panels show all the groups of distances to center from a control or a pulsed fly.

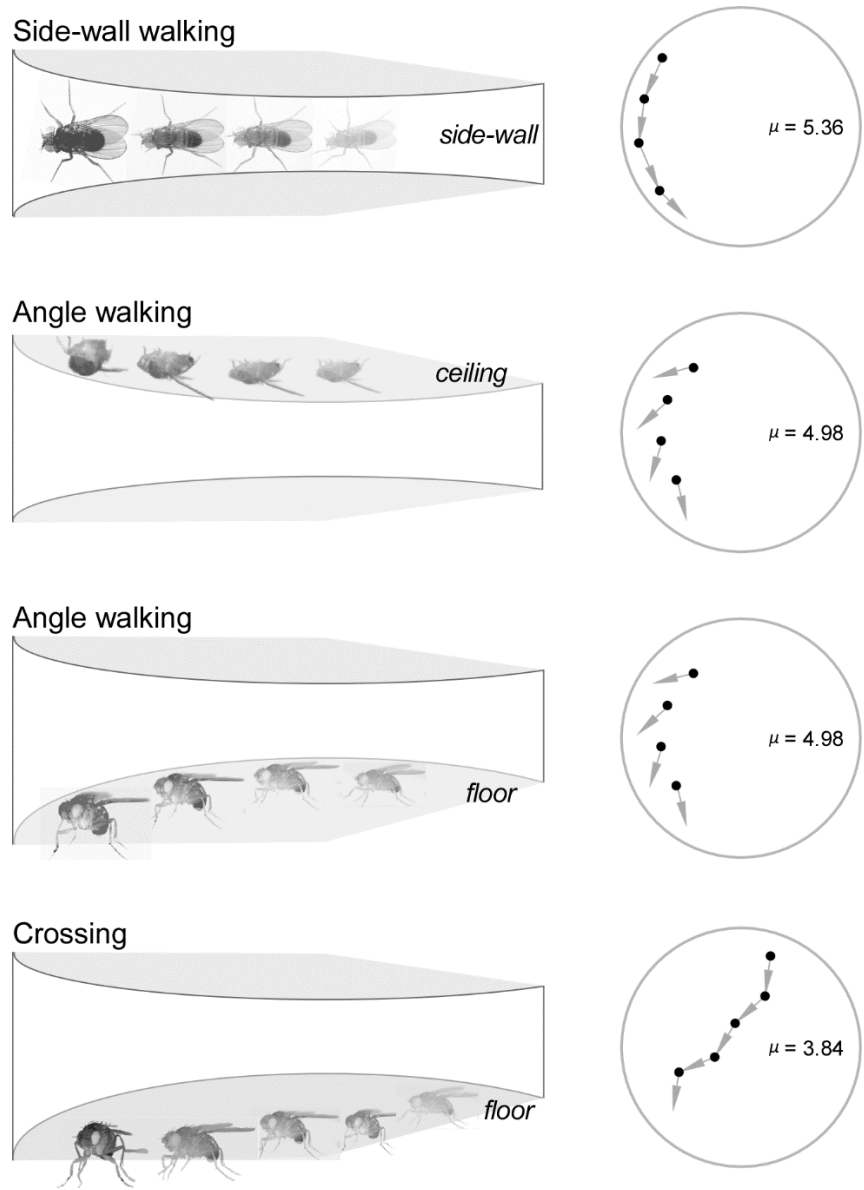

Figure S3: Schematic illustrations of walking structures in a circular arena. There is only one fly in each arena. Left panels: 3D demonstrations of walking structures (including crossing, side-wall walking, and angle walking) and relevant fly locations in the arena. The side-wall, ceiling, and floor are labeled. Right panels: 2D views of fly positions (black dots) and their instant directions (grey arrows) while walking. Values of location parameter $\mu$ are provided (see text for further interpretation). Grey circle indicates the edge of the arena. During angle walking, locations on the ceiling appear to be the same as those on the floor in the 2D view.